\documentclass[preprint,aps,pra,showpacs,floatfix]{revtex4-1}
\usepackage[T2A,T1]{fontenc}
\usepackage[utf8]{inputenc}
\renewcommand{\vec}[1]{{\mbox{\boldmath$#1$}}}

\usepackage{graphicx}
\usepackage{nicefrac}
\usepackage{amsmath}
\usepackage{amsfonts}
\usepackage{amssymb}
\usepackage{amsthm}
\usepackage{epsf}
\usepackage{bm}
\usepackage{multirow}
\usepackage{siunitx}
\usepackage{mathtools}
\usepackage{booktabs}
\newcommand{\be}{\begin{eqnarray}}
\newcommand{\ee}{\end{eqnarray}}

\usepackage{color}
\definecolor{BLUE}{rgb}{0.0,0.0,1.0}

\usepackage{soul}
\usepackage{soulutf8}
\soulregister\cite7
\soulregister\ref7

\begin{document}

\title{Relativistic calculations of the chemical properties of the superheavy element with $Z=119$ and its homologues}

\author{I.~I.~Tupitsyn, A.~V.~Malyshev, D.~A.~Glazov, M.~Y.~Kaygorodov,
Y.~S.~Kozhedub, I.~M.~Savelyev, V.~M.~Shabaev}

\affiliation{St.~Petersburg State University, 199034 St.~Petersburg, Russia }

\begin{abstract}
Relativistic calculations of the electronic structure of the superheavy element of the eighth period --- eka-francium ($Z=119$) and its homologues, which form the group of alkali metals, are performed in the framework of the configuration-interaction method and many-body perturbation theory using the basis of the Dirac--Fock--Sturm orbitals (DFS). The obtained values of the ionization potentials, electron affinities, and root-mean-square radii are compared with the corresponding values calculated within the non-relativistic approximation. A comparison with the available experimental data and the results of previous theoretical calculations is given as well. The analysis of the obtained results indicates a significant influence of the relativistic effects for the francium and eka-francium atoms, which leads to a violation of the monotonic behaviour of the listed above chemical properties as a function of the alkaline-element atomic number. In addition, the quantum electrodynamics corrections to the ionization potentials are evaluated by employing the model Lamb-shift operator (QEDMOD).
\end{abstract}


\maketitle

%
%
\section{Introduction}
%
%
The synthesis and study of superheavy nuclei represent one of the most important tasks of nuclear physics~\cite{Oganessian:2016:901, Oganessian:2019:5, Nazarewicz:2018:537}. The physical and chemical properties of the corresponding superheavy elements (SHE) are of both applied and fundamental interest. On the one hand, the very process of detecting SHE atoms is closely related to their ability to form compounds. On the other hand, the question of the extent to which the properties of lighter homologues are preserved for SHE, in fact, determines the limits of applicability of the Mendeleev's periodic law. At the same time, the experimental study of the properties of SHE is currently difficult due to the extremely small number of synthesized atoms and their very short lifetimes. Therefore, theoretical data on the properties of SHE are in great demand. Obviously, the corresponding calculations, along with the electron--electron correlations, must take into account the relativistic and quantum electrodynamics (QED) effects. The electronic structure of SHE is intensively studied by several scientific groups, see, e.g., Refs.~\cite{Kaldor_2003, Schadel_2014, Eliav_2015, Schwerdtfeger_2015, Jerabek:2018:053001, Lackenby:2018:042512, Eliav_2019, Pershina_2019, Kaygorodov:2020:036} and references therein.

This paper presents calculations of the chemical properties of eka-francium ($Z=119$) and its lighter homologues, which form the first group of the periodic table, the group of alkali metals. Previously, in various works (see, e.g., \cite{Nefedov_2006} and references therein), the main configuration of the eka-francium was established to be [Rn]$5f^{14}\,6d^{10}\,7s^2\,7p^6\,8s^1$. The most accurate relativistic calculations of the electronic structure of eka-francium and its homologues have been performed by the coupled-cluster method~\cite{Landau_2001, Eliav_2005_IP, Eliav_2005_EA, Lim_2005, Borschevsky_2013}. In the present work, we employ the relativistic configuration-interaction method and many-body perturbation theory (CI+MBPT) in the basis of Dirac--Fock--Sturm (DFS) orbitals. The contributions of the relativistic effects are found by comparing the results of the relativistic calculations with the data obtained by the same methods in the non-relativistic limit by scaling the speed of light. The contributions of the QED effects are evaluated within the model Lamb-shift-operator approach. It is shown that, despite the significant contribution of the relativistic effects, the properties of eka-francium correspond to the first group of the periodic table.
%
%
\section{Theory. Brief description of the calculation method}
%
%
In the present work, we employ the relativistic Dirac--Coulomb--Breit (DCB) Hamiltonian~\cite{Sucher_1980, Mittleman_1981}
\begin{equation}
\label{eq:H_DCB}
  \hat H_{\rm DCB} = \Lambda_{+} \, \left[ \hat H_{\rm D} + \hat H_{\rm C} + \hat H_{\rm B} \right ]  \Lambda_{+} 
\,.
\end{equation}
$\hat H_{\rm D}$ is the sum of the one-electron four-component Dirac Hamiltonians,
\begin{equation}
  \hat H_{\rm D} = \sum_{i}^{N} \, \hat h_{\rm D}(i) 
\,,
\end{equation}
\begin{equation}
  \hat{h}_{\rm D}(i) = -ic \big(\vec{\alpha}_{i} \cdot \vec{\nabla}_{i}\big) + c^2 (\beta_{i}-1) + V_{\rm n}(r_i) 
\,,
\end{equation}
where $N$ is the number of electrons, $c$ is the speed of light, $\vec{\alpha}$ and $\beta$ are the Dirac matrices, $V_{\rm n}(r)$ is the nuclear potential. Here and below in the work, the atomic units are used. To describe the nuclear-charge distribution, we use the Fermi model with the thickness parameter equal to $t=2.3~{\rm fm}$. The value of the root-mean-square nuclear radius for eka-francium is calculated employing the empirical formula $\langle r^2\rangle^{1/2}=[0.77 A^{1/3}+0.98]~{\rm fm}$, where $A$ is the atomic number which is assumed to be 295 for this element. For homologues, the values of the root-mean-square nuclear radii are taken from the Ref.~\cite{Angeli:2013:69}. The operator $\hat H_{\rm C}$ is the two-electron Coulomb-interaction operator,
\begin{equation}
  \hat H_{\rm C} = \frac{1}{2} \, \sum_{i \ne j}^{N} \, \frac{1}{r_{ij}} 
\,, \qquad 
  r_{ij} = |\vec{r}_i-\vec{r}_j| 
\,.
\end{equation}
The operator $\hat H_{\rm B}$ is the two-electron frequency-independent Breit operator in the Coulomb gauge,
\begin{equation}
\label{breit}
  \hat H_{\rm B} \,=\, - \frac{1}{2} \, \sum_{i \ne j}^{N} \,\frac{1}{r_{ij}} \bigg[ \big(\vec{\alpha}_{i} \cdot  \vec{\alpha}_{j}\big) 
           + \frac{1}{2} \big( \vec{\alpha}_{i} \cdot \vec{\nabla}_{i} \big) 
             \big( \vec{\alpha}_{j} \cdot \vec{\nabla}_{j} \big) \, r_{ij} \bigg] 
\,.
\end{equation}
The operator $\Lambda_{+}$ projects many-electron wave functions onto the space of the Slater determinants made up of the one-electron positive-energy states.
%
%
\subsection{The Dirac--Fock--Sturm method}
%
%
Within the DCB approximation, the many-electron wave function $\Psi(J M)$ with the quantum numbers $J$ and $M$ can be obtained in the form of an expansion in terms of the configuration-state functions (CSF) $\Phi_{I} (JM)$:
\begin{equation}
\label{CI_eq1}
  \Psi(JM) = \sum_{I} \,C^{JM}_{I} \, \Phi_{I} (JM) 
\,, 
\end{equation}
where $\Phi_{I}(JM)$ are the eigenfunctions of the operators $\hat J^2$ and $\hat J_z$, and are constructed as a linear combinations of the Slater determinants of a given relativistic configuration.

The Ritz variational principle in the CSF space reduces the solution of the DCB equation to the eigenvalue problem for the Hamiltonian matrix:
\begin{equation}
\label{CI_eq2}
  \sum_{K} H_{KI} \, C_K^{JM} = E_I(J) \, C_I^{JM} 
\,,
\end{equation}
where $H_{KI}=\langle \Phi_K\mid \hat H_{\rm DCB} \mid \Phi_I \rangle$, and the indices $I$ and $K$ enumerate the different CSFs.

In the present paper, the one-electron wave functions $\psi_i$ are obtained by the Dirac--Fock--Rutan method in the basis of Dirac--Fock--Sturm (DFS) orbitals $\varphi_k$:
\begin{equation}
\label{DFR}
  \psi_i = \sum_{k} \, u_{ki} \, \varphi_k 
\,.
\end{equation}
In turn, the DFS basis is constructed as follows. The one-electron wave functions, obtained by numerically solving the Dirac--Fock (DF) integro-differential equations \cite{Bratzev_1977}, are employed as the orbitals $\varphi_k$, which are occupied in the ground and low-lying excited states. For virtual (high-lying vacant) one-electron states, the functions~$\varphi_k$ are obtained by numerically solving the Dirac--Fock--Sturm equations \cite{Tupitsyn_2003,Tupitsyn_2005},
\begin{equation}
\label{DFS}
  \left[ \hat h_{\rm DF} - \varepsilon_0 \right ] \, \varphi_k = \mu_k \, W(r) \, \varphi_k 
\,,
\end{equation}
where $\hat h_{\rm DF}$ is the Dirac--Fock operator, $\varepsilon_0$ is the reference one-electron energy, and $W(r)$ is a positive weight function tending to zero at infinity. We note that all the DFS orbitals have approximately the same characteristic size and the same asymptotics at infinity, determined by the reference energy $\varepsilon_0$:
\begin{equation}
  \varphi_k(r) \xrightarrow[r\to\infty]{} C_k \, \exp(-\sqrt{2\varepsilon_0} \, r) 
\,.
\end{equation}
As the weight function $W(r)$, the function tending to a constant for $r \to 0$ is chosen:
\begin{equation}
  W(r) = \frac{1-\exp(-(\alpha  r)^2)}{(\alpha r)^2} 
\,.
\end{equation}
We note that in the present study the same basis of the one-electron functions~$\psi_{i}$ is used in all the calculations.
%
%
\subsection{Configuration-interaction method in the combination with many-body perturbation theory (CI+MBPT)}
%
%
To account for the electron--electron correlations, in the present work a combination of the configuration-interaction (CI) method and many-body perturbation theory (MBPT) is employed. To construct the configuration space, the concept of the restricted active space (RAS)~\cite{Olsen_1988} in the basis of the DFS orbitals~\cite{Tupitsyn_2003, Tupitsyn_2005} is used. According to this approach, the set of the one-electron functions is divided into 4 subgroups: RAS0, RAS1, RAS2, and RAS3.

The RAS0 subspace includes the so-called frozen core orbitals, excitations from which are not taken into account. The frozen-core electrons create the one-particle Coulomb and exchange potentials of the core and do not participate in the CI$+$MBPT calculation. The subspace RAS1 includes the orbitals of the outer core, the single and double excitations from which are taken into account by means of perturbation theory. The active valence orbitals are assigned to the RAS2 subspace, and the virtual highly excited orbitals are assigned to RAS3. In this paper, the single and double excitations from RAS1 and RAS2 are taken into account by the perturbation theory up to the second order. In the RAS2 valence space, the CI function is constructed by solving the eigenvalue problem for the Hamiltonian matrix~(\ref{CI_eq2}).

Perturbation theory is well suited to account for a large number of excitations, and it also has the important property of dimensional consistency. The configuration-interaction method allows one to treat the excitations within the active space to all orders. The approach we employ flexibly and effectively uses the advantages of both methods, which becomes especially important with an increase in the number of electrons.
%
%
\subsection{Quantum electrodynamics corrections}
%
In the present work, to account for the QED effects we employ the one-electron model Lamb-shift operator $\hat{h}^{\rm QED}$ (QEDMOD), which was proposed in Ref.~\cite{Shabaev_2013} and is widely used in various quantum-chemistry calculations, see, e.g., Refs.~\cite{Tupitsyn_2016, Pasteka:2017:023002, Machado:2018:032517, Zaytsev:2019:052504, Kumar:2020:012503, Shabaev_2020}. The QED correction is evaluated as the difference of the two total energies calculated by the CI method. In one of the calculations, the model QED operator is included into the DCB Hamiltonian, while the other calculation is carried out without the model operator. The model operator $\hat{h}^{\rm QED}$ consists of two parts,
\begin{equation}
\label{qed}
  \hat{h}^{\rm QED} = \hat{h}^{\rm VP} +  \hat{h}^{\rm SE} 
\,.
\end{equation}
The operators $\hat h^{\rm VP}$ and $\hat{h}^{\rm SE}$ represent the contributions of the vacuum polarization and the self energy, respectively. The operator $\hat{h}^{\rm VP}$ can be represented as a sum of the local Uehling potential $V_{\rm Uehl}(r)$ and the Wichmann--Kroll potential $V_{\rm WK}(r)$
\begin{equation}
  \hat{h}^{\rm VP} = V_{\rm Uehl}(r) + V_{\rm WK}(r) 
\,.
\end{equation}
The Uehling potential represents the leading contribution in terms of the powers of interaction with the nucleus~\cite{Serber:1935:49, Uehling:1935:55}. It can be obtained both numerically and analytically using approximate formulas from Ref.~\cite{Fullerton_76}. Evaluation of the Wichmann-Kroll potential is a much more complicated problem~\cite{Soff_88, Manakov_89, Perrson_93}. However, the contribution of this potential can be estimated with a fairly high accuracy by employing the approximate formulas presented in Ref.~\cite{Fainshtein_91}.

The self-energy operator $\hat{h}^{\rm SE}$, which is described in detail in Ref.~\cite{Shabaev_2013}, is modeled as a sum of the local short-range potential $V_{\rm loc}^{\rm SE}$ and the non-local separable (finite-dimensional) potential,
\begin{equation}
  \hat{h}^{\rm SE}=V_{\rm loc}^{\rm SE} + \sum\limits_{i,k=1}^n | \phi_i \rangle B_{ik} \langle \phi_k | 
\,.
\end{equation}
Here $\phi_i$ are the projection functions selected in a specific way, and the matrix $B_{ik}$ is defined to reproduce exactly the known diagonal and non-diagonal matrix elements of the one-loop self-energy operator in the basis of hydrogen-like functions~\cite{Shabaev_2013}. The computer code for generating the QEDMOD operator is presented in Refs.~\cite{Shabaev_2015, Shabaev_2018}.

The simplest approach to evaluate the QED corrections is to average the model operator $\hat{h}^{\rm QED}$ with the many-electron wave functions. In our calculations, we include the model operator $\hat{h}^{\rm QED}$ into the DCB Hamiltonian from the very beginning.  This approach additionally takes into account higher orders in the model potential, which can give a noticeable contribution~\cite{Shabaev_2020}.
%
%
\section{Results and discussions}
%
%
In this paper, the calculations of eka-francium ($Z=119$) and its homologues are performed using the relativistic DF and CI+MBPT methods in the basis of Dirac--Fock--Sturm orbitals. In order to study the change in chemical properties in the transition from light alkali metals to superheavy alkali metals, we have calculated the ionization potentials, the electron affinities, and the characteristic sizes of the valence $ns$-shell.

\begin{figure}
\begin{center}
\includegraphics[width=0.65\linewidth]{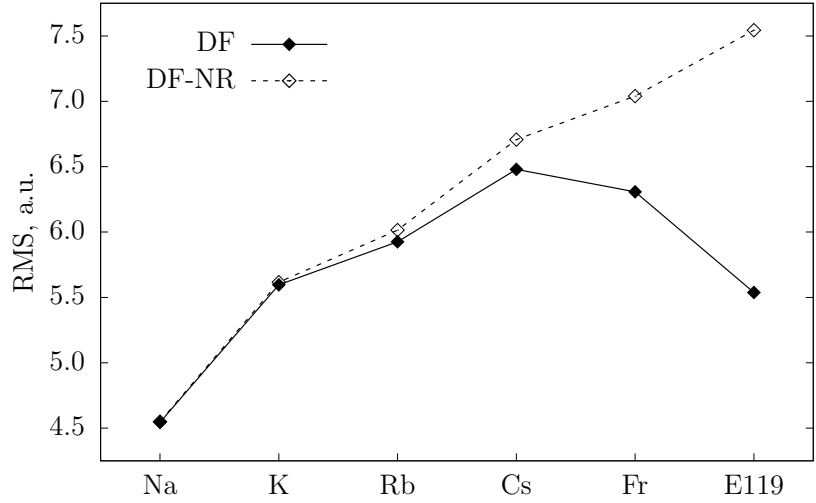}
\caption{\label{fig:RMS}
Root-mean-square radii (RMS) of the valence $ns$-shell (in a.u.).
DF~corresponds to the relativistic Dirac--Fock method, DF-NR~indicates the non-relativistic limit.
}
\end{center}
\end{figure}

\begin{figure}
\begin{center}
\includegraphics[width=0.65\linewidth]{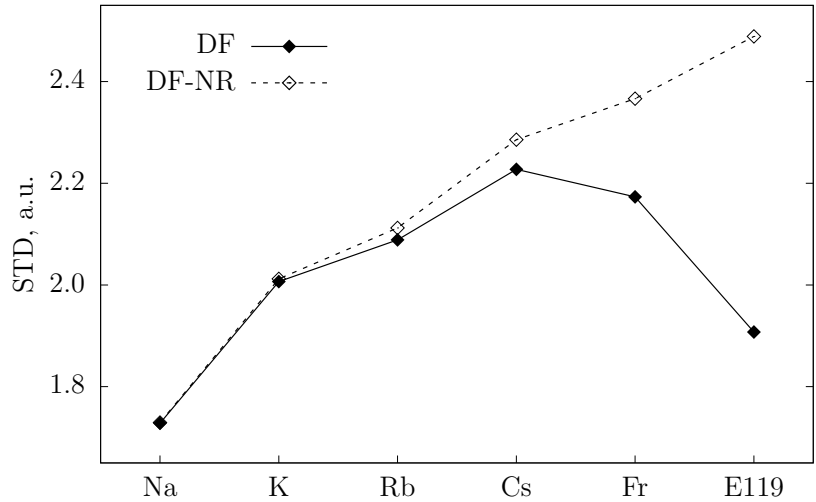}
\caption{\label{fig:STD}
Standard deviations (STD) for the valence $ns$-shell (in a.u.).
Notations are the same as in Fig.~\ref{fig:RMS}.
}
\end{center}
\end{figure}

Table~\ref{RMS_tab} shows the relativistic and non-relativistic values of the root-mean-square radii (${\rm RMS}=\langle ns | r^2 | ns \rangle^{1/2}$) and the standard deviations (${\rm STD}=\langle ns | (r-\overline r)^2 | ns \rangle^{1/2}$), which characterize the size of the alkali-metal atoms and the width of the electron-density distribution of the valence $ns$-shell, respectively. In addition, the same data are presented in Figs.~\ref{fig:RMS} and \ref{fig:STD}. We note that the value of STD is a measure of the localization of the valence $ns$-shell. The relativistic values of RMS and STD are obtained by the Dirac--Fock method, and their non-relativistic analogues are calculated by increasing the speed of light by a factor of 100.
The table and figures clearly show that the non-relativistic values of RMS and STD increase monotonically with the growth of $Z$, which indicates the delocalization of the valence $ns$-states. However, as it can be seen from the same figures, taking into account the relativistic effects leads to a violation of the monotonic behaviour of the RMS and STD values, starting from $Z=87$. This fact is related to the contraction of the $ns$-orbitals and to the increase in the degree of their localization, which is a purely relativistic effect.
%
\begin{table}[t]
\centering

\renewcommand{\arraystretch}{0.9}
 
\caption{\label{RMS_tab} 
         Root-mean-square radii (RMS) and standard deviations (STD) 
         for the valence $ns$-shell (in a.u.). 
         DF~corresponds to the relativistic Dirac--Fock method,
         DF-NR~indicates the non-relativistic limit.
         }

\vspace*{2mm}
\begin{tabular}{
                @{\,\,\,}c@{\quad}|
                S[table-format=4.4]|
                S[table-format=3.4]|
                S[table-format=4.4]|
                S[table-format=3.3]
               }
               
\hline

   
   \multicolumn{1}{c|}{\multirow{2}{*}{$Z$}}     &
   \multicolumn{2}{c|}{\rule{0pt}{1.2em}RMS}     &
   \multicolumn{2}{c}{STD}                        \\
   
   \cline{2-5}  
   
                                                  &
   \multicolumn{1}{c|}{DF-NR}                      &
   \multicolumn{1}{c|}{DF}                        &
   \multicolumn{1}{c|}{DF-NR}                      &
   \multicolumn{1}{c}{DF}                         \\
   
\hline   
                       
 \rule{0pt}{3.6ex}  11 &   4.55  &   4.54  &   1.73  &   1.73   \\ 
                    19 &   5.62  &   5.60  &   2.01  &   2.01   \\ 
                    37 &   6.01  &   5.93  &   2.11  &   2.09   \\ 
                    55 &   6.71  &   6.48  &   2.29  &   2.23   \\ 
                    87 &   7.04  &   6.31  &   2.37  &   2.17   \\ 
                   119 &   7.54  &   5.54  &   2.49  &   1.91   \\[1mm] 

\hline

\end{tabular}%

\end{table}
%
%
%
Tables \ref{IP_tab} and \ref{EA_tab} present the ionization potentials (IP) and electron affinities (EA) of eka-francium and its lighter homologues, evaluated by means of the relativistic methods DF and CI$+$MBPT, as well as their values in the non-relativistic limit. For all the elements, the basis consists of the Dirac--Fock orbitals for the atomic and ionic shells occupied in the ground state and a set of virtual Dirac--Fock--Sturm orbitals. The basis of the virtual orbitals includes 7 $s$, 7 $p$, 7 $d$, 5 $f$, 4 $g$, and  3 $h$  Sturm functions. When calculating the ionization potentials, we use the basis constructed for a neutral atom. In the calculations of the electron affinities, the basis constructed for the negative ion is employed.
%
\begin{table}[t]
\centering

\renewcommand{\arraystretch}{0.9}

\caption{\label{IP_tab} 
         Ionization potentials (IP) of the elements from the first group of the periodic table (in eV).
         DF~corresponds to the Dirac--Fock method,
         CI+MBPT~stands for the configuration-interaction method combined with many-body perturbation theory;
         NR~indicates the non-relativistic limit.
         }

\vspace*{2mm} 
\begin{tabular}{
                @{\,\,\,}c@{\quad}|
                S[table-format=6.6]|
                S[table-format=4.5]|
                S[table-format=4.5]|
                S[table-format=3.4]
               }
               
\hline

   \multicolumn{1}{c|}{\rule{0pt}{1.2em}$Z$}    &
   \multicolumn{1}{c|}{~DF-NR~}                     &
   \multicolumn{1}{c|}{DF}                        &
   \multicolumn{1}{c|}{CI$+$MBPT-NR}             &
   \multicolumn{1}{c}{~CI$+$MBPT}                \\      
   
\hline   
                       
 \rule{0pt}{3.6ex}  11 &  4.961  &  4.961  &  5.151  &  5.159   \\ 
                    19 &  4.012  &  4.026  &  4.303  &  4.323   \\ 
                    37 &  3.808  &  3.808  &  4.097  &  4.177   \\ 
                    55 &  3.365  &  3.486  &  3.723  &  3.889   \\ 
                    87 &  3.208  &  3.603  &  3.578  &  4.079   \\ 
                   119 &  2.995  &  4.327  &  3.258  &  4.780   \\[1mm] 

\hline

\end{tabular}%

\end{table}
\begin{figure}
\begin{center}
\includegraphics[width=0.65\linewidth]{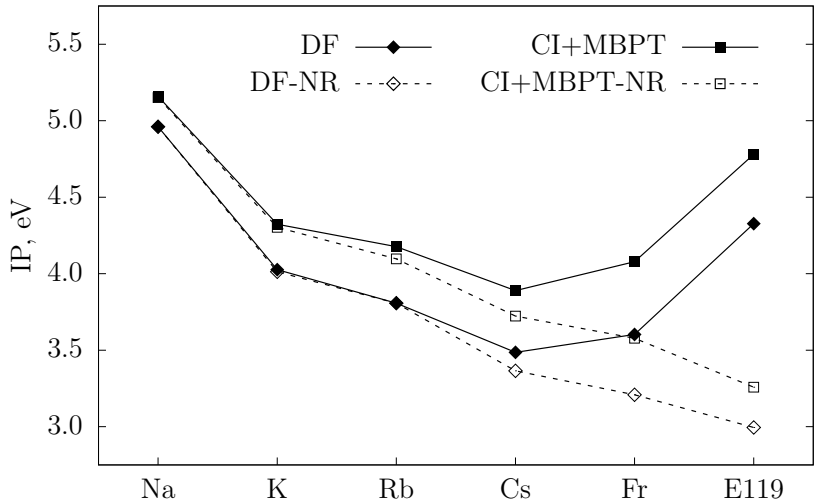}
\caption{\label{fig:IP}
Ionization potentials (IP) of the elements from the first group of the periodic table (in eV). 
DF~corresponds to the Dirac--Fock method, 
CI+MBPT~stands for the configuration-interaction method combined with many-body perturbation theory;
NR~indicates the non-relativistic limit.
}
\end{center}
\end{figure}
%
As it can be seen from Tables \ref{IP_tab} and \ref{EA_tab}, and Figs.~\ref{fig:IP} and \ref{fig:EA}, the values of the ionization potentials and electron affinities evaluated in the non-relativistic limit by the DF and CI$+$MBPT methods decrease monotonically with increasing the atomic number. This behaviour of the IP and EA values correlates with the increase in the degree of delocalization of the non-relativistic RMS and STD values. Accounting for the relativistic effects violates the monotonic behaviour of the IP and EA values, starting from $Z=87$, as it takes place for the RMS and STD values.
\begin{table}[t]
\centering

\renewcommand{\arraystretch}{0.9}

\caption{\label{EA_tab} 
         Electron affinities (EA) of the elements from the first group of the periodic table (in eV).
         Notations are the same as in Table~\ref{IP_tab}.
         }

\vspace*{2mm} 
\begin{tabular}{
                @{\,\,\,}c@{\quad}|
                S[table-format=6.6]|
                S[table-format=4.5]|
                S[table-format=4.5]|
                S[table-format=3.4]
               }
               
\hline

   \multicolumn{1}{c|}{\rule{0pt}{1.2em}$Z$}    &
   \multicolumn{1}{c|}{~DF-NR~}                     &
   \multicolumn{1}{c|}{DF}                        &
   \multicolumn{1}{c|}{CI$+$MBPT-NR}             &
   \multicolumn{1}{c}{~CI$+$MBPT}                \\      
   
\hline   
                       
 \rule{0pt}{3.6ex}  11 &  0.363  &  0.364  &  0.551  &  0.553   \\ 
                    19 &  0.281  &  0.282  &  0.504  &  0.505   \\ 
                    37 &  0.259  &  0.264  &  0.473  &  0.488   \\ 
                    55 &  0.227  &  0.237  &  0.453  &  0.475   \\ 
                    87 &  0.214  &  0.251  &  0.431  &  0.498   \\ 
                   119 &  0.197  &  0.388  &  0.423  &  0.674   \\[1mm] 

\hline

\end{tabular}%

\end{table}
%
\begin{figure}
\begin{center}
\includegraphics[width=0.65\linewidth]{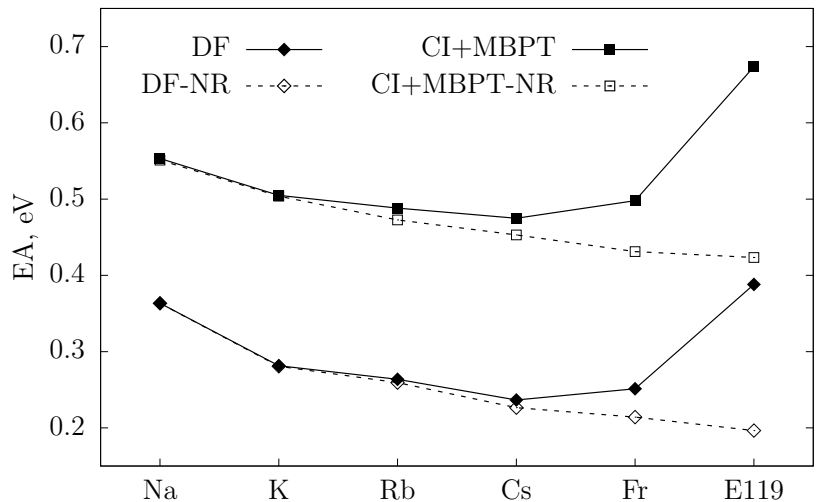}
\caption{\label{fig:EA}
Electron affinities (EA) of the elements from the first group of the periodic table (in eV). 
Notations are the same as in Fig.~\ref{fig:IP}.
}
\end{center}
\end{figure}
Comparing the values obtained by the DF and CI+MBPT methods, we can judge the contribution of the correlation effects to the IP and EA values. For the ionization potentials, the correlation energy is of the order of 0.3--0.4 eV for all the elements of the group of alkaline elements presented in Table~\ref{IP_tab}, except for Na ($Z=11$), for which it is approximately 0.2 eV. The contribution of the correlation energy to the electron affinities lies in the range 0.2-0.3 eV, and it increases monotonically with the growth of $Z$.
%
\begin{table}[t]
\centering

\renewcommand{\arraystretch}{0.9}

\caption{\label{IP_QED_tab} 
         QED contributions to the ionization potentials of the elements from the first group of the periodic table (in meV).
         QEDMOD~corresponds to the model Lamb-shift operator,
         VRAD~stands for the radiation potential;
         DF~designates the~Dirac--Fock method,
         CI+MBPT~indicates the configuration-interaction method combined with many-body perturbation theory.
         }

\vspace*{2mm} 
\begin{tabular}{
                @{\,\,\,}c@{\quad}|
                S[table-format=-5.3]@{\,\,\,}|
                S[table-format=-4.3]|
                S[table-format=-4.3]@{\,\,\,}|
                S[table-format=-4.2]
               }
               
\hline
   
   \multicolumn{1}{c|}{\multirow{2}{*}{$Z$}}     &
   \multicolumn{2}{c|}{QEDMOD}                     &
   \multicolumn{1}{c|}{VRAD}                       &
   \multicolumn{1}{c}{Ref.}                     \\
   
   \cline{2-4}
   
                                                  &
   \multicolumn{1}{c|}{DF}                         &
   \multicolumn{1}{c|}{CI$+$MBPT}                &
   \multicolumn{1}{c|}{DF}                         &   
   \multicolumn{1}{c}{\cite{Eliav_2005_IP}} \\
   
\hline   
                       
 \rule{0pt}{3.6ex}  11 &   -0.3  &   -0.3  &   -0.3  &          \\ 
                    19 &   -0.5  &   -0.6  &   -0.5  &          \\ 
                    37 &   -1.3  &   -1.5  &   -1.3  &   -1.3   \\ 
                    55 &   -2.0  &   -2.3  &   -2.0  &   -2.2   \\ 
                    87 &   -4.8  &   -5.1  &   -4.7  &   -3.6   \\ 
                   119 &  -12.3  &  -12.6  &   -9.8  &  -10.3   \\[1mm] 

\hline

\end{tabular}%

\end{table}
Table~\ref{IP_QED_tab} presents the results of our calculations of the QED corrections to the ionization potentials. As it can be seen from the table, there are minimal differences between our data obtained by the DF and CI$+$MBPT methods with the model QED potential (QEDMOD) \cite{Shabaev_2013}. For comparison, the fourth column (VRAD) presents our results obtained by the DF method using the radiation potential~\cite{Flambaum_2005}. The last column shows the corresponding results from Ref.~\cite{Eliav_2005_IP}.
%
\begin{table}[t]
\centering

\renewcommand{\arraystretch}{0.9}

\caption{\label{IP_tot_tab} 
         Ionization potentials with the QED corrections included for the elements from the first group 
         of the periodic table (in eV). 
         Comparison with the experimental data and results from Ref.~\cite{Eliav_2005_IP}.
         }

\vspace*{2mm} 
\begin{tabular}{
                @{\,\,\,}c@{\quad}|
                S[table-format=1.3]|
                S[table-format=1.3]|
                S[table-format=1.3]
               }
               
\hline
   
   \multicolumn{1}{c|}{\rule{0pt}{1.2em}$Z$}       &
   \multicolumn{1}{c|}{~~~This work~~}                 &
   \multicolumn{1}{c|}{~~Ref.~\cite{Eliav_2005_IP}~~}    &
   \multicolumn{1}{c}{~~Experiment~~}                \\  
   
\hline   
                       
 \rule{0pt}{3.6ex}  11 &  5.159  &         &  5.139   \\ 
                    19 &  4.323  &         &  4.341   \\ 
                    37 &  4.175  &  4.181  &  4.177   \\ 
                    55 &  3.887  &  3.901  &  3.894   \\ 
                    87 &  4.074  &  4.079  &  4.073   \\ 
                   119 &  4.768  &  4.783  &          \\[1mm] 

\hline

\end{tabular}%

\vspace*{2mm}

\begin{flushleft}
\textit{Comment.} Experimental data are taken from 
Refs.~\cite{Ciocca:1992:4720, Baugh:1998:1585} for Na; 
Refs.~\cite{Lorenzen:1981:370, Lorenzen:1983:300} for K;
Refs.~\cite{Lorenzen:1983:300, Johansson:1961:135} for Rb;
Ref.~\cite{Deiglmayr:2016:013424} for Cs;
Ref.~\cite{Arnold:1990:3511} for Fr.
\end{flushleft}

\end{table}
\begin{table}[t]
\centering

\renewcommand{\arraystretch}{0.9}

\caption{\label{EA_tot_tab} 
         Electron affinities of the elements from the first group of the periodic table (in eV).        
         Comparison with the experimental data and results from Ref.~\cite{Eliav_2005_EA}.
         }

\vspace*{2mm} 
\begin{tabular}{
                @{\,\,\,}c@{\quad}|
                S[table-format=1.3]|
                S[table-format=1.3]|
                S[table-format=1.3]
               }
               
\hline
   
   \multicolumn{1}{c|}{\rule{0pt}{1.2em}$Z$}       &
   \multicolumn{1}{c|}{~~~This work~~}                 &
   \multicolumn{1}{c|}{~~Ref.~\cite{Eliav_2005_EA}~~}    &
   \multicolumn{1}{c}{~~Experiment~~}                \\  
   
\hline   
                       
 \rule{0pt}{3.6ex}  11 &  0.553  &  0.548  &  0.548   \\ 
                    19 &  0.505  &  0.503  &  0.501   \\ 
                    37 &  0.488  &  0.486  &  0.486   \\ 
                    55 &  0.475  &  0.471  &  0.472   \\ 
                    87 &  0.498  &  0.486  &          \\ 
                   119 &  0.674  &  0.649  &          \\[1mm] 

\hline

\end{tabular}%

\vspace*{2mm}

\begin{flushleft}
\textit{Comment.} Experimental data are taken from 
Ref.~\cite{Hotop:1985:731} for Na; 
Refs.~\cite{Slater:1978:201, Andersson:2000:022503} for K;
Ref.~\cite{Frey:1978:L589} for Rb;
Refs.~\cite{Slater:1978:201, Scheer:1998:684} for Cs.
\end{flushleft}

\end{table}

%
Table~\ref{IP_tot_tab} gives the values of the ionization potentials obtained in the present work by the CI$+$MBPT method taking into account the QED corrections. This table shows also the results of the IP calculations by the coupled-cluster method~\cite{Eliav_2005_IP} with the QED corrections taken into account as well, and the available experimental values. As it can be seen from the comparison, our data agree well with the results of Ref. \cite{Eliav_2005_IP} and with the experiment within the uncertainty of our calculations, which we estimate as 0.1 eV. This uncertainty estimate is due to the relatively small size of the basis and the contribution of the triple excitations, which are not taken into account in this work.

The values of the electron affinities (without taking into account the QED corrections)  obtained in the present work by the CI$+$MBPT method are presented in the Table~\ref{EA_tot_tab}. This table shows also the results of the calculations employing the coupled-cluster method \cite{Eliav_2005_EA}. There is also a good agreement between our data and the results of Ref.~\cite{Eliav_2005_EA} and the experiment. We estimate the uncertainty of our EA calculations to be 0.1 eV as well.

%
%
\section{Conclusion}
%
%
In this paper, the chemical properties of superheavy element eka-francium ($Z=119$) and its lighter homologues were considered. The root-mean-square radii and standard deviations of the valence $ns$-shell, which characterize the degree of the localization of the valence electron density, were calculated. The obtained results indicate that the structure of the valence shells is preserved in the transition from lighter homologues to eka-francium. This shows that it, as expected, belongs to the first group of elements of the periodic table.

We also obtained the non-relativistic and relativistic values of the ionization potentials and electron affinities for eka-francium and its homologues. It was found that the relativistic effects violate the monotonic decrease of IP and EA taking place in the non-relativistic limit. This fact is explained by the effect of contracting the $s$- and $p$-shells, when the relativistic corrections are taken into account, and it is consistent with a similar statement formulated earlier, e.g., in Ref.~\cite{Eliav_2019}.

In the future, we plan to extend the present CI+MBPT calculations to a number of superheavy elements with $Z=118$--$130$.
%
%
\section{Funding}
%
%
The reported study was funded by RFBR and ROSATOM, project number 20-21-00098.
%
\section{Conflict of interest}
%
The authors declare that they have no conflict of interest.
%
%

\end{document}